# High-throughput screening of spin Hall conductivity in 2D materials

Fu Li[1], Xiaoxiong Liu[2], Vikrant Chaudhary[1,3], Ruiwen Xie[1], Chen Shen[1*], Hao Wang[1*] and Hongbin Zhang[1]

[1]Institute of Materials Science, Technology University of Darmstadt, 64287, Darmstadt, Germany

[2]Department of Physics, Southern University of Science and Technology, 518055, Shenzhen, China

[3]Department of Physics and CSMB, Humboldt University of Berlin, 12489, Berlin, Germany

Corresponding authors e-mail: chenshen@tmm.tu-darmstadt.de;

haowang@tmm.tu-darmstadt.de

**Abstract**

Two-dimensional (2D) materials with large spin Hall effect (SHE) have attracted significant attention due to their potential applications in next-generation spintronic devices. In this work, we perform high-throughput (HTP) calculations to obtain the spin Hall conductivity (SHC) of 4486 non-magnetic compounds in 2Dmatpedia database and identify six materials with SHC exceeding 500 ($\hbar$/e)(S/cm), surpassing those of previous known materials. Detailed analysis reveals that the significant SHC can be attributed to spin-orbit coupling (SOC)-induced gap openings at Dirac-like band crossings. Additionally, the presence of mirror symmetry further enhances the SHC. Beyond the high SHC materials, 57 topological insulators with quantized SHCs have also been identified. Our work enables rapid screening and paves the way for experimental validation, potentially accelerating the discovery of novel 2D materials optimized for spintronics applications.

**Introduction**

Spintronics, which utilizes the spin degree of freedom in addition to charge for information processing and storage, has long been a vibrant field of research and continues to offer promising opportunities for next-generation electronics[1]. It offers several advantages, including higher data storage density, lower power consumption, and faster processing speed[2]. One of the key mechanisms underlying spintronics is SHE, in which an external electric field induces a transverse spin current[3,4]. SHE generally comprises three contributions: an intrinsic part[5], associated with the quantum geometry of the electronic structure (i.e., spin Berry curvature), and two extrinsic parts namely skew scattering and side-jump[6]. In this work, we consider only the intrinsic SHC. From a computational point of view, the (inverse) SHC quantitatively characterizes the SHE, with larger SHC values indicating higher charge-to-spin conversion efficiency and improved device performance, as demonstrated both experimentally and theoretically[7-9]. Moreover, the intrinsic SHE originates from SOC, without requiring an external magnetic field or intrinsic magnetization[10-12], thereby enabling device miniaturization and integration.

Since the discovery of graphene[13], the emergent 2D materials have opened new avenues for spintronic applications. Their atomically thin nature offers high tunability and the potential for integration with flexible substrate[14-17]. A wide range of novel physical phenomena have been observed in 2D materials, including the quantum Hall effect[18], topological semimetals[19], Klein tunneling[20], the valley Hall effect[21], and topological insulators[22]. In particular, the 2D quantum SHE (QSHE), observed in 2D topological insulators (TIs), is widely regarded as a hallmark of topological phases. They feature nontrivial electronic properties protected by time-reversal symmetry and is characterized by a nonzero $\mathbb{Z}_2$ topological invariant[23, 24]. For instance, graphene is essentially a 2D TI with quantized SHC[25]; however, its band gap is too small for practical applications. Therefore, it is crucial to explore other 2D materials as potential TIs or with significant SHC. In this regard, 2D compounds comprising heavy elements are of particular interest as they can exhibit large SHC due to enhanced SOC, such as monolayer $Ta_4Se_2$ and $Y_2Br_2$[26].

However, traditional trial-and-error approaches severely limit the efficiency of discovering new spintronic materials, since large-scale experimental screening for high-performance candidates is impractical. Consequently, HTP density functional theory (DFT) calculations have become essential for identifying materials with enhanced transport properties. Notable efforts include the work of Noky et al.[27] and our previous studies[28-30], which investigated intrinsic anomalous Hall conductivity and anomalous Nernst conductivity in ferromagnetic materials. Similarly, Zhang et al.[31] and Ji et al.[32] conducted high-precision, HTP ab-initio calculations of intrinsic SHC in three-dimensional crystals. Despite these advances, many promising material candidates and their underlying intrinsic mechanisms remain to be further explored. In particular, understanding the interplay between symmetry, topology, and the SHE in 2D materials remains an open and compelling challenge.

In this work, we employ an HTP workflow to compute the SHC of 4486 non-magnetic 2D materials. We identify 27 candidate materials exhibiting SHC values exceeding 300 ($\hbar$/e)($S/cm$). Detailed analysis of one representative material, $Te_3Os_4$ reveals that its pronounced SHC originates from strong SOC, which is similar to PW, where multiple Dirac-like band crossings lead to an SOC-induced band gap opening, resulting in a large SHC. In addition, we identify 57 novel 2D TIs with quantized SHC. Our study not only introduces an efficient framework for computing SHC across extensive 2D material databases but also sheds light on the underlying physical mechanisms responsible for high SHC.

**Computational method**

We developed a workflow to evaluate the intrinsic topological transport properties based on automated construction of maximally localized Wannier functions (MLWFs)[33]. The workflow begins with crystal structures collected from the 2Dmatpedia database[34] which contains 6351 materials. The first-principles calculations were performed using the VASP code, in which the projected augmented wave (PAW) method is implemented[35]. The exchange-correlation functional was using the generalized gradient approximation (GGA) as parameterized by Perdew, Burke, and Ernzerhof (PBE)[36], and the SOC effect was considered. The k-mesh used for the calculations was Γ-centred, with a density of 0.02 Å$^{-1}$. The cut-off energy for the plane waves was set to 550 eV. After performing DFT calculations, we excluded all 1865 magnetic 2D materials, resulting in 4486 non-magnetic candidates. Subsequently, the MLWFs were constructed using an in-house developed automatic wannierization workflow that integrates VASP with Wannier90[37]. The intrinsic SHC was evaluated using Wannierberri based on the tight-binding model Hamiltonian constructed from MLWFs[38]. A denser k mesh of 400 × 400 × 1 was employed to perform the Brillouin zone (BZ) integration for the intrinsic SHC, and adaptive k-mesh refinement was used to deal with the rapid variation of the spin Berry curvature in the BZ. According to the Kubo formula, the intrinsic SHC can be written as:

$$\sigma_{xy}^z = -\frac{e^2}{\hbar}\frac{1}{VN_k^3}\sum_k \Omega_{xy}^z(\mathbf{k})$$

where the k-resolved term reads:

$$\Omega_{xy}^z(\mathbf{k}) = \sum_n f_{nk}\Omega_{n,xy}^z(\mathbf{k}),$$

and the band-projected spin Berry curvature (SBC) can be defined as:

$$\Omega_{n,xy}^z(\mathbf{k}) = \hbar^2 \sum_{m\neq n} \frac{-2\text{Im}\left[\langle n\mathbf{k}|\frac{1}{2}\{\hat{\sigma}_z,\hat{v}_x\}|m\mathbf{k}\rangle\langle m\mathbf{k}|\hat{v}_y|n\mathbf{k}\rangle\right]}{(\epsilon_{nk}-\epsilon_{mk})^2},$$

where $|n\mathbf{k}\rangle/|m\mathbf{k}\rangle$ and $\epsilon_{n/mk}$ are the Bloch wave functions and eigenenergies of bands n and m, respectively. $\frac{1}{2}\{\hat{\sigma}_z,\hat{v}_x\}$ describes a spin current operator, where $\hat{\sigma}_z$ is the Pauli matrix, and $\hat{v}_{x/y}$ are the velocity operators.

**Results**

The distribution of the calculated absolute SHC for the full set of materials is shown in Fig. 1(a). The SHC values exhibit an approximately linear decay on a logarithmic scale, indicating that the majority of materials fall within the low-SHC regime, typically below 200 ($\hbar$/e)(S/cm). Among the 4486 non-magnetic materials analyzed, we identified 27 candidates with SHC values exceeding 300 ($\hbar$/e)(S/cm), and 6 of these exceed 500 ($\hbar$/e)(S/cm). Our analysis reveals that in most high-SHC materials, the large conductivity originates from high crystal symmetry combined with the presence of heavy-atoms systems, which giving rise to multiple Dirac-like crossings in the electronic structure in the absence of SOC. When SOC is introduced, these Dirac-like crossings are split or gapped out, generating significant SBC near the Fermi level, which in turn results in a large SHC [39-41]. We further observe that materials with mirror symmetry tend to exhibit greater SHC. Among all materials with SHC exceeding 100 ($\hbar$/e)(S/cm), 71.6% possess at least three mirror planes. This proportion increases to 76.8% for materials with SHC above 200 ($\hbar$/e)(S/cm). To further illustrate the influence of mirror planes on SHC, Fig. 1(b) shows the distribution of SHC across point groups, categorized by crystal system. From this distribution, it is evident that most high-performance SHC compounds originate from the monoclinic, trigonal, and tetragonal crystal systems. We also observe that materials with point groups such as 3, −3, 4, −4, 32, and −6 tend to exhibit weaker SHE, likely due to the lack of mirror symmetry. Notably, after excluding compounds where strong SOC is the dominant factor of SHC, the −3m, 3m, and 4/mmm point groups emerge as the ones most consistently associated with higher SHC values. These results highlight the crucial role of mirror symmetry in enhancing SHC[31] (see Fig. S1 and Table S1).

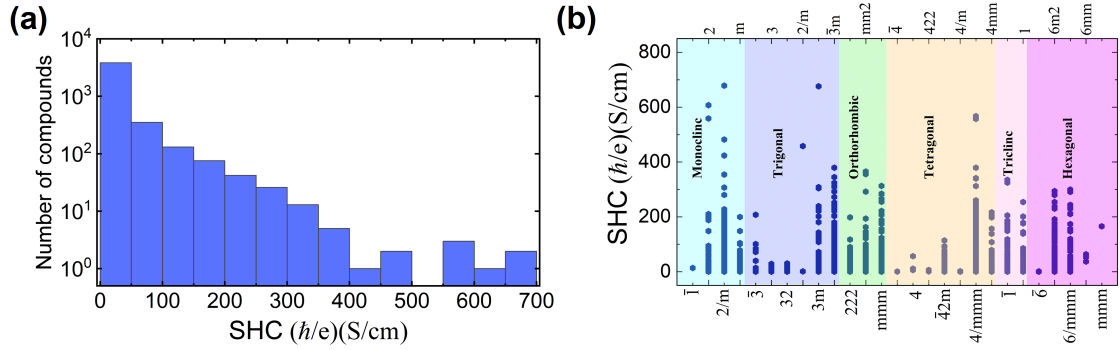

Fig.1 (a) Distribution of SHC values for all calculated materials. (b) Classification by point groups for calculated materials.

Table 1 lists 20 2D materials exhibiting the highest SHC at the Fermi level. Among the listed candidates, $Bi_4Pb_3$ demonstrates the highest SHC value of 678 $(\hbar/e)(S/cm)$, followed by $Te_4Os_3$ (676 $(\hbar/e)(S/cm)$) and $Ge_3Bi_4$ (607 $(\hbar/e)(S/cm)$). These values are significantly higher than those of experimentally realized excellent performance transition metal dichalcogenides, such as $WTe_2$[42, 43] with 20–300 $(\hbar/e)(S/cm)$ and 2D van der Waals heterostructures[44, 45]. It is worth noting that among the top 20 candidates, several materials—such as $Bi_4Pb_3$, $Sn_3Bi_4$, and $Te_3Os_4$—exhibit large SHC primarily due to their strong SOC (the SOC strength dependent SHC of $Bi_4Pb_3$ is illustrated in Fig. S2). In contrast, $Ta_2S$, PW, $CaPb_2$, and other materials, achieve large SHC through a distinct mechanism: SOC-induced gap openings at band crossings around the Fermi energy (the band structures are shown in Fig. S3 and S4), which generate substantial SBC. Moreover, the maxima of SHC can be tailored by shifting the Fermi energy level via a combination of strain and doping[46]. Therefore, identifying materials with large SHC values close to the Fermi level is particularly important. Table 1 reveals compounds with maximal SHC ($SHC_{max}$) and the corresponding shifts relative to the Fermi level. For example, $Ta_2S$ exhibits an $SHC_{max}$ of −739 $(\hbar/e)(S/cm)$, which is located 0.13 eV above the Fermi level (see Fig. S5). This means that a slight adjustment of the Fermi level can significantly enhance the SHC in experiments.

Table.1 List of top 20 2D materials with promising transport properties, displaying lattice constants, space group (SG), SHC values (at Fermi-level), $SHC_{max}$ within [−0.40, 0.40] eV with respect to the Fermi-level (ΔE)

| Compounds | a Å | b Å | SG | SHC $(\hbar/e)(S/cm)$ | $SHC_{max}[\Delta E]$ $(\hbar/e)(S/cm)[eV]$ |
|---|---|---|---|---|---|
| $Bi_4Pb_3$ | 5.10 | 5.10 | C2 | 678 | 799 (0.32) |
| $Te_4Os_3$ | 3.55 | 3.55 | P-6m2 | 676 | 677 (0.01) |
| $Ge_3Bi_4$ | 4.50 | 4.50 | C2 | 607 | 667 (0.36) |
| $Ta_2S$ | 3.33 | 3.33 | P4/nmm | -567 | -739 (0.13) |
| $Sn_3Bi_4$ | 4.54 | 4.54 | C2 | 558 | 579 (-0.11) |
| $Tl_3Os_2$ | 3.34 | 3.34 | P4/mmm | 557 | 562 (0.02) |
| PW | 3.13 | 3.13 | C2/m | -482 | -485 (-0.01) |
| Pd | 2.70 | 4.66 | P-3m1 | 458 | 650 (-0.15) |
| Pt | 2.69 | 4.67 | C2/m | 423 | 502 (-0.16) |
| $HPb_3$ | 4.97 | 4.97 | P-3m1 | 379 | 379 (0.00) |
| $Tl_2Os$ | 2.95 | 2.95 | P4/mmm | 379 | 616 (-0.34) |
| GePt | 3.88 | 2.94 | Pmn | 365 | 368 (-0.04) |
| BeBi | 3.39 | 3.45 | Pmm2 | 356 | 375 (-0.05) |
| Pb | 4.89 | 4.89 | C2/m | 354 | 354 (0.00) |
| CaPb$_2$ | 3.53 | 3.53 | P-3m1 | 344 | 346 (0.03) |
| IrC$_3$ | 4.63 | 4.63 | P4/mbm | 341 | 430 (-0.12) |
| PtPb$_3$ | 6.67 | 6.67 | P4/mbm | 340 | 509 (0.20) |
| Te$_2$Ru | 6.55 | 6.53 | Pcca | 331 | 390 (0.12) |
| TiP$_3$ | 7.64 | 7.64 | Pcca | 325 | 340 (-0.05) |
| Ce$_2$Mg | 3.15 | 3.15 | P-3m1 | 323 | 372 (0.05) |

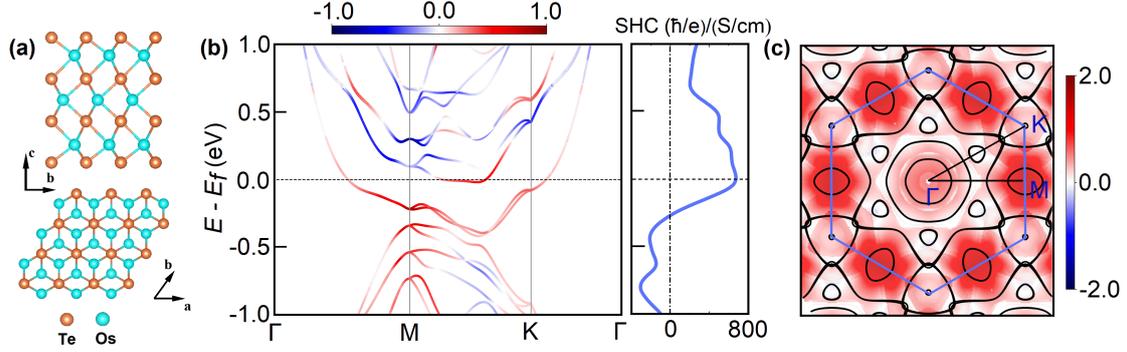

Fig.2 (a) Crystal structure of Te$_3$Os$_4$, the orange and cyan balls stand for Te and Os atoms, respectively. (b) Band structure projected by SBC (left panel) and SHC as a function of energy (right panel) of Te$_3$Os$_4$. (c) The k-resolved SBC in 2D BZ of Te$_3$Os$_4$.

To further investigate the origin of the large SHC, we selected Te$_3$Os$_4$ for a detailed analysis, as shown in Fig.2. We examine both its electronic structure and the SBC distribution within the whole BZ. Figure 2(a) presents the crystal structure of Te$_3$Os$_4$, which belongs to the space group P-6m2 and the point group 3m. From the SBC resolved band structure of Te$_3$Os$_4$ shown in Fig. 2(b), the positive SBC contributions are predominantly concentrated around the M point in the valence bands, whereas negative contributions are mainly associated with the conduction bands. Notably, the significant band splitting along the M–K path suggests strong SOC effects (see Fig. S6). As a result, Te$_3$Os$_4$ exhibits a sizable SHC near the Fermi level, approximately 676 ($\hbar/e$)($S/cm$). The k-resolved SBC for all the valence bands in reciprocal space also confirms this observation as demonstrated in Fig. 2(c). It is evident that the majority area of the 2D BZ is dominated by the positive contributions, especially around the M point, while the SBC near the Γ point is close to zero as the conduction and valence bands are widely separated in energy.

In addition, to explain another mechanism that can improve SHC, we select PW as a representative example, as shown in Fig. 3. The large SHC is closely associated with the band crossings that exist in the absence of SOC. Without necessary symmetry protection, band crossings open a gap when SOC is introduced, leading to a large SHC. The electronic band structures of PW without and with SOC are shown in Fig. 3(a) and Fig. 3(b), respectively. In the absence of SOC, multiple band crossings are generated near the Fermi level, highlighted by red circles. As expected, with the inclusion of SOC, these bands become gapped (the 3D bands shown in Fig. S7) and the Fermi level lies almost within those gaps. Such band degeneracy lifting produced a large SBC, as seen in the red circles in Fig. 3(b), where the cyan and red lines represent negative and positive SBC, respectively. A prominent peak in the SHC appears near the Fermi level corresponding to these regions of large SBC, as shown in Fig. 3(c). This results in SBC

"hotspots", as indicated by the intense blue areas in Fig. 3(e), precisely where the original crossings occurred along the Γ-Y and Γ-M paths. In order to verify the total SHC mainly arise from the hotspots, we calculated the total SHC and the SHC contributed by the eight selected band-crossing points, shown as red and blue lines in Fig. 3(f). The contribution of each point is evaluated by calculating the SHC within a square area of side length 0.15, centered at the crossing point. The eight squares cover merely ~18% of the BZ, yet their SHC contribution (–350 (ℏ/e)(S/cm)) amounts to 68.8% of the total at the Fermi level (–509 (ℏ/e)(S/cm)). These results confirm that band crossings make substantial contributions to the overall SHC in PW.

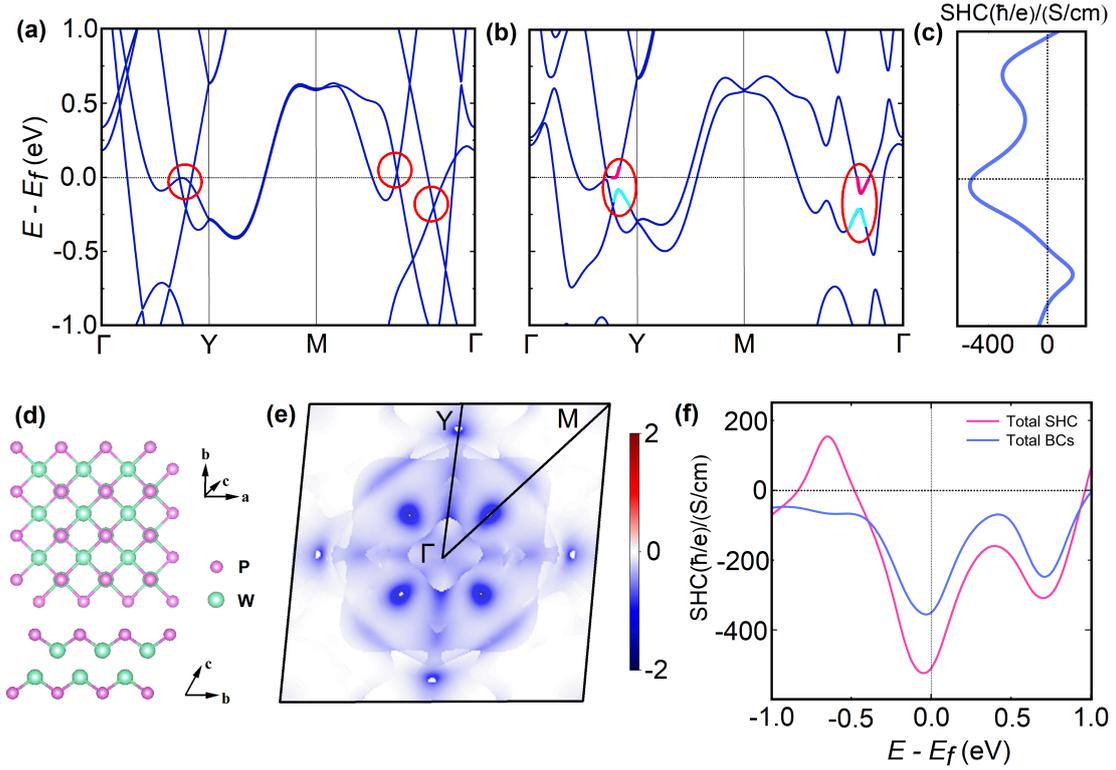

Fig.3 Electronic band structure of PW, (a) without and (b) with SOC, respectively. Dirac points and gapped points are highlight on the k-path. (c) SHC as a function of energy of PW. (d) Crystal structure of PW, the magenta and light green balls stand for P and W atoms, respectively. (e) The k-resolved spin Berry curvature in a slice of 2D BZ of PW. (f) The total SHC (red) and the contribution from all eight crossing points to SHC (blue) as a function of energy.

While heavy metallic 2D materials offer valuable insights for future spintronic applications, insulating systems are equally important. In particular, we demonstrate that quantized SHC serves as a key signature of TI phases in 2D systems. Firstly, we present a crystal system classification of screened TIs to analyze their structural characteristics, which can be categorized into hexagonal, tetragonal, triclinic, orthorhombic, monoclinic, and trigonal systems, as summarized in Table 2. According to our HTP screening, we identified 57 2D TIs, with their structural information and band structures provided in the supplementary information. Among them, the trigonal structure is the most prevalent, accounting for 17 compounds. To the best of our knowledge, the 2D compounds previously reported in the literature are marked with superscript references, while the others are new candidates identified in this work, as shown in Table 2. In the literature, R. R. Freitas et al. reported that the band gap values of $InBi_3$, $GaBi_3$, and $AlBi_3$ are strongly influenced by SOC and by the nature of group-III elements, leading to distinct topological properties[57]. Similarly, M. Hirayama et al. proposed HfBr electrides as a novel platform for topological materials[53].

Table.2 Screened topological insulators with different space group

| System | Compounds |
|---|---|
| Hexagonal | AuI, LiTl, $Ag_2Te$[47] |
| Tetragonal | NaMgAs, NaMgSb, $SrSbSe_2F$ |
| Triclinic | $Ta_2S_3$, $Ta_2Se_3$[19], $V_2Se_3$, $SbBr_5$, $PdI_2$, $PtI_2$, BiSb, $Ca_2Ge$, $TaI_5$ |
| Orthorhombic | $YbBi_2$, $ZrTe_5$[48, 49], $HfTe_5$[48], $TiTe_5$, $ZrSe_5$[50], $ZrS_5$, $HfSe_5$[51], $TiSe_5$, $HfS_5$[51], $BiBr_5$ |
| Monoclinic | HfBr[52, 53], $EuP_2$, $EuBi_2$, CrP, $Cu_2Te$[47], MoAs, $NbTe_4Ir$[45], $EuAs_2$, $Ta2Se_3$, BiSe[54], BiTe[55], $Tl_2Te_3$, Pb, $Ta_2Se_3$ |
| Trigonal | HgO, CuF, AuF, AuBr, AuCl, HgS, $Sn_3H$, $InBi_3$[56], $GaBi_3$[56], $BSb_3$, $AlBi_3$[56], HgTe[57], GePb, SnPb[58], TlSe[59], $HPb_3$, $Sb_3C$, $NbI_3$ |

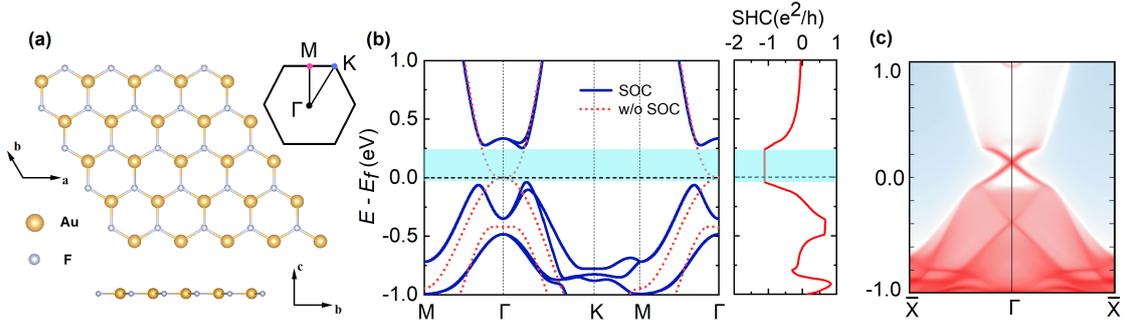

Fig 4 (a) Crystal structure and BZ of AuF, the golden and silver balls stand for Au and F atoms, respectively. (b) The band structure with (solid blue line) and without (red dotted line) SOC, and the QSHC of AuF. (c) The calculated edge states of AuF.

Among the screened 2D TIs, we highlight several representative materials to illustrate the relationship between crystal structure and topological properties. As the first example, AuF is presented in Fig. 4, where panel (a) shows its crystal structure with a honeycomb-like arrangement. This structural similarity to graphene suggests the possibility of topological behavior. In Fig. 4(b), we present the orbital-resolved band structures both without and with SOC. In the absence of SOC, the minority and majority spin bands are degenerate. These degenerate states split and open a band gap of 0.31 eV when SOC is introduced, indicating a hallmark of topological insulators. To confirm the nontrivial nature of this SOC-induced band gap, we calculated the $\sigma_{xy}^{z}$ as a function of energy relative to the Fermi level. The quantized plateau observed within the SOC gap indicates the realization of the quantum spin Hall phase. Furthermore, the nontrivial phase is also confirmed by the existence of gapless edge states. We used the iterative method to compute the edge Green's function for a semi-infinite system, and the corresponding local density of states is presented in Fig. 4(c). A pair of gapless edge states can be clearly observed within the SOC-induced gap, further validating the topological nature of AuF.

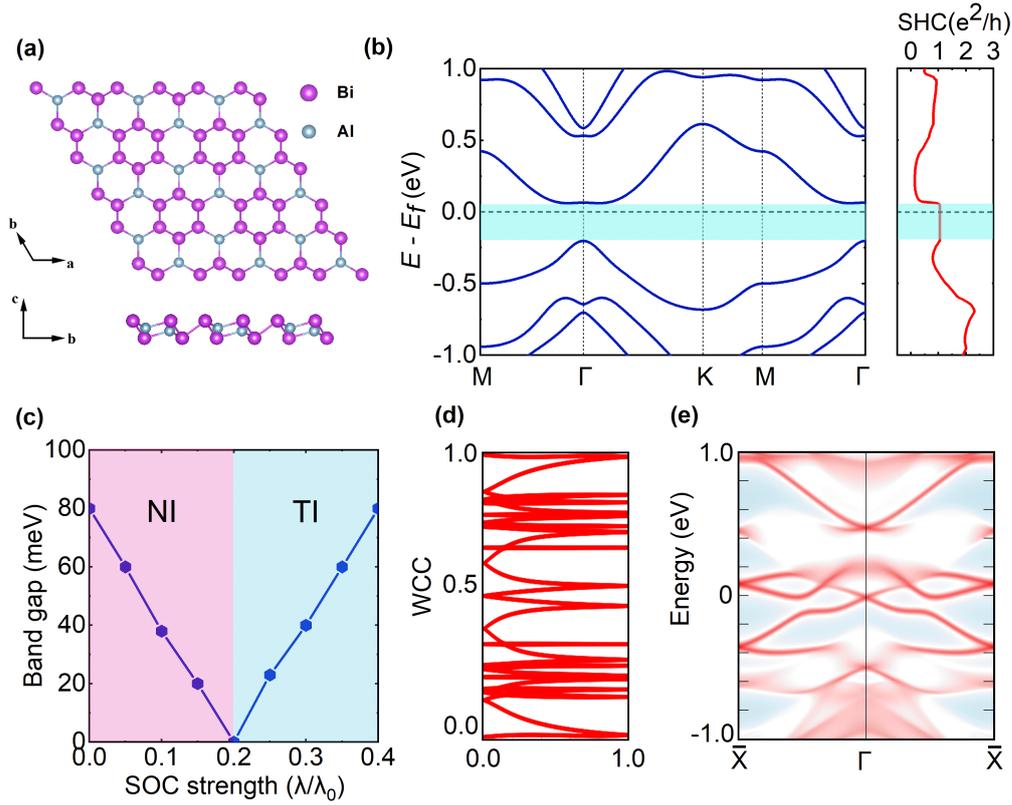

Fig 5 (a) Crystal structure of AlBi$_3$, the purple and light steel blue balls stand for Bi and Al atoms, respectively. (b) The band structure with SOC, and the quantum SHC of AlBi$_3$. (c) The band gap of AlBi$_3$ as a function of SOC strength. (d) The evolution of WCCs for AlBi$_3$. (e) The calculated edge state of AlBi$_3$

AlBi$_3$ is another case with nontrivial topological features. Figure 5(a) displays its crystal structure, which consists of a buckled honeycomb lattice. In order to explore its topological properties, we present the electronic structure and SHC in Fig. 5(b), which exhibits a direct bandgap of 0.27 eV at the Γ point and a quantized SHC of $\sigma_{xy}^{z} \approx e^2/h$ within the gap. Additionally, we examined the effect of SOC strengths on the AlBi$_3$ system. As shown in Fig. 5(c), in the absence of SOC, the system shows a direct bandgap of 0.08 eV. The bandgap gradually decreases until the SOC strength increases to 0.2 $\frac{\lambda}{\lambda_0}$, indicating that the system remains a trivial insulator. However, when the SOC strength exceeds 0.2 $\frac{\lambda}{\lambda_0}$, a band inversion occurs, signifying the emergence of a topological insulating phase. This nontrivial character is further confirmed by the evolution of the Wannier charge centers (WCCs) and by the presence of edge states. As shown in Fig. 5(d), the WCCs exhibit an odd number of crossings with an arbitrary horizontal line, while Fig. 5(e) reveals the presence of gapless helical edge states.

**Summary**

In conclusion, we carried out HTP first-principles screening of 4486 non-magnetic 2D materials from the 2Dmatpedia database, identifying candidates with high SHC and topological insulating behavior. In total, 27 materials exhibit SHC values exceeding 300 ($\hbar/e$)($S/cm$). Beyond the well-known role of strong SOC, we emphasize that materials with multiple mirror symmetries tend to exhibit larger SHC, with those possessing three or more mirror planes showing particularly high SHC on average. In addition, we uncovered candidates governed by different mechanisms. For example, in PW the enhanced SHC originates from Dirac-like band crossings near the Fermi level, which open band gaps under SOC and generate substantial SBC. Furthermore, we identified 57 2D TIs based on the presence of quantized SHC, AuF demonstrates quantized SHC and gapless helical edge states, confirming that it is a quantum spin Hall insulator. Similarly, $AlBi_3$ exhibits a SOC-driven topological phase transition, which can be verified through band-gap evolution, WCC crossings, and the emergence of helical edge states. Overall, our HTP framework provides an efficient strategy for discovering novel 2D materials with desirable spintronic properties. The identified candidates offer promising platforms for next-generation spintronic devices, quantum computing, and low-power electronic applications.

## Acknowledgements

The authors gratefully acknowledge the computing time provided to them on the high-performance computer Lichtenberg at the NHR Centers NHR4CES at TU Darmstadt. This project was supported by the China Scholarship Council.